\documentclass[
reprint,
superscriptaddress,
showpacs,
amsmath,amssymb,
aps,prl,
]{revtex4-1}
\usepackage{amssymb}
\usepackage{amsmath}
\usepackage{booktabs}
\usepackage{tabularx}
\usepackage{braket}
\usepackage{float}
\usepackage{graphicx}
\usepackage{dcolumn}
\usepackage{bm}
\usepackage{subfigure}
\usepackage{color}

\usepackage{epstopdf, epsfig}
\usepackage{blindtext}
\usepackage{longtable}
\usepackage{amsmath}
\usepackage{mathtools}
\usepackage{enumitem}
\usepackage[T1]{fontenc}
\usepackage{mathptmx}

\usepackage[colorlinks,citecolor = blue, linkcolor=red,hyperindex,CJKbookmarks]{hyperref}

\begin{document}

\title{Periodically modulated thermal convection}

\author{Rui Yang}
\affiliation{Physics of Fluids Group, Max Planck Center for Complex Fluid Dynamics, MESA+ Institute and J.M.Burgers Center for Fluid Dynamics, University of Twente, P.O. Box 217, 7500 AE Enschede, The Netherlands}
\affiliation{Max Planck Institute for Dynamics and Self-Organisation, Am Fassberg 17, 37077 G{\"o}ttingen, Germany}
\author{Kai Leong Chong}
\affiliation{Physics of Fluids Group, Max Planck Center for Complex Fluid Dynamics, MESA+ Institute and J.M.Burgers Center for Fluid Dynamics, University of Twente, P.O. Box 217, 7500 AE Enschede, The Netherlands}
\author{Qi Wang}
\affiliation{Physics of Fluids Group, Max Planck Center for Complex Fluid Dynamics, MESA+ Institute and J.M.Burgers Center for Fluid Dynamics, University of Twente, P.O. Box 217, 7500 AE Enschede, The Netherlands}
\affiliation{Department of Modern Mechanics, University of Science and Technology of China, Hefei 230027, China}
\author{Roberto Verzicco}
\affiliation{Physics of Fluids Group, Max Planck Center for Complex Fluid Dynamics, MESA+ Institute and J.M.Burgers Center for Fluid Dynamics, University of Twente, P.O. Box 217, 7500 AE Enschede, The Netherlands}
\affiliation{Dipartimento di Ingegneria Industriale, University of Rome 'Tor Vergata', Via del Politecnico 1, Roma 00133, Italy}
\affiliation{Gran Sasso Science Institute - Viale F. Crispi, 7, 67100 L'Aquila, Italy}
\author{Olga Shishkina}
\affiliation{Max Planck Institute for Dynamics and Self-Organisation, Am Fassberg 17, 37077 G{\"o}ttingen, Germany}
\author{Detlef Lohse}\email{d.lohse@utwente.nl}
\affiliation{Physics of Fluids Group, Max Planck Center for Complex Fluid Dynamics, MESA+ Institute and J.M.Burgers Center for Fluid Dynamics, University of Twente, P.O. Box 217, 7500 AE Enschede, The Netherlands}
\affiliation{Max Planck Institute for Dynamics and Self-Organisation, Am Fassberg 17, 37077 G{\"o}ttingen, Germany}

\date{\today}

\begin{abstract}
Many natural and industrial turbulent flows are subjected to time-dependent boundary conditions. Despite being ubiquitous, the influence of temporal modulations (with frequency $f$) on global transport properties has hardly been studied. Here, we perform numerical simulations of Rayleigh-B\'enard (RB) convection with time periodic modulation in the temperature boundary condition and report how this modulation can lead to a significant heat flux (Nusselt number $\rm{Nu}$) enhancement. Using the concept of Stokes thermal boundary layer, we can explain the onset frequency of the Nu enhancement and the optimal frequency at which Nu is maximal, and how they depend on the Rayleigh number Ra and Prandtl number $\rm{Pr}$. From this, we construct a phase diagram in the 3D parameter space ($f$, $\rm{Ra}$, $\rm{Pr}$) and identify: (i) a regime where the modulation is too fast to affect $\rm{Nu}$; (ii) a moderate modulation regime, where $\rm{Nu}$ increases with decreasing $f$ and (iii) slow modulation regime, where $\rm{Nu}$ decreases with further decreasing $f$. Our findings provide a framework to study other types of turbulent flows with time-dependent forcing.
\end{abstract}

\maketitle
Turbulent flows driven by time-dependent forcing are common in  nature and industrial applications \cite{Berger1972periodic,Davis1976}. For example, the Earth's atmosphere circulation is driven by periodical heating from solar radiation, the ocean tidal current by periodical gravitational attractions from both the Moon and the Sun, and the blood circulation by the beating heart.

The focus of earlier work was on the response amplitude and phase delay. In periodically driven turbulence in shear flows, a mean-field theory has been used to analyze the resonance maxima of the Reynolds number \cite{Lohse2000, VonderHeydt2003a}. Periodic forcing in other turbulent systems, for example, in the homogeneous isotropic turbulence \cite{VonderHeydt2003b,Kuczaj2006}, pipe flow \cite{jelly2020direct,Cheng2020forcing,Papadopoulos2016pulsating,He2009}, channel flow \cite{Weng2016numerical,Bhaganagar2008}, Taylor-Couette flow \cite{Verschoof2018a,Barenghi1989}, and Rayleigh-B\'enard convection \cite{Jin2008,Sterl2016,Niemela2008}, is also shown to have highly non-trivial response properties.

Here we picked turbulent Rayleigh-B\'enard convection \cite{ahlers2009,Lohse2010,Chilla2012} as model system to study how modulation influences global heat transport. In this system different modulation methods have been studied previously, such as bottom temperature modulation \cite{Jin2008,Niemela2008}, rotation modulation \cite{Sterl2016,Geurts2014} and gravity modulation \cite{Rogers2000,Gresho1970effects}. Intuitively, we may expect that the effect on time-averaged global quantities is limited because the net force averaged over a cycle vanishes. Indeed, with bottom temperature modulation in experiments, only a small enhancement ($\approx 7\%$) of the heat flux has been observed so far \cite{Jin2008,Niemela2008}. However, the effects of modulation in temperature have not yet been fully explored because of the experimental challenge in having a broad range of modulation frequency due to thermal inertia.

In this Letter, we numerically study modulated RB convection within a wide range (more than four orders of magnitude) of modulation frequency at the bottom plate temperature and observe a significant ($\approx 25\%$) enhancement in heat transport. To explain our findings, we show the relevance of the Stokes thermal boundary layer (BL), which is analogous to the classical one for an oscillating plate \cite{Kundu2001}, in determining the transitional frequency for the heat transport enhancement and the optimal frequency for the maximal heat transport. In particular, we calculate the transition between the different regimes in phase space and show how they depend on the Rayleigh and Prandtl numbers. Our modulation method is complementary to hitherto used concepts of using additional body force or modifying the spatial structure of the system to enhance heat transport, for example, adding surface roughness \cite{Jiang2018,Ciliberto1999,emran2020natural}, shaking the convection cell \cite{chao2020}, including additional stabilizing forces through geometrical modification, rotation, inclination, or a second stabilizing scalar field, etc \cite{Stevens2009,Yang2015,Chong2015,zwirner2020influence,Chong2017}.

RB convection is the flow in a container heated from below and cooled from above. Next to the aspect ratio of the horizontal and vertical extensions of the container, the dimensionless control parameters are the Rayleigh number $\rm{Ra}=\alpha gH^3 \Delta/(\nu\kappa)$ and the Prandtl number $\rm{Pr}=\nu/\kappa$, with $\alpha$, $\nu$, and $\kappa$ being, respectively, the thermal expansion coefficient, kinematic viscosity and thermal diffusivity of the fluid, $g$ the gravitational acceleration and $\Delta$ the temperature difference between the bottom and top boundaries. The time, length and temperature are made dimensionless by the free-fall time $\tau=\sqrt{H/\alpha g\Delta}$, the height $H$ of the container, the temperature difference $\Delta$, respectively. In the following, all quantities are dimensionless, if not otherwise explicitly stated. In the periodically modulated RB, we give a sinusoidal modulation signal to the bottom temperature as
\begin{equation}
    \theta_{bot}=1+A\cos(2\pi ft).
\end{equation}
For modulated RB, two more parameters have to be introduced, namely the modulation frequency $f$ and its amplitude $A$, which is kept fixed in this study, $A = 1$. The efficiency of the heat transport and flow strength in the system are represented in terms of the Nusselt number $\rm{Nu}$ (the dimensionless heat flux) and the Reynolds number $\rm{Re}$. Direct Numerical Simulation (DNS) for incompressible Oberbeck-Bousinesq flow are employed \cite{verzicco1996}; the numerical details are provided in the Supplemental Material. The DNS are conducted in a two-dimensional square box with no-slip and impermeable boundary conditions (BCs) for all walls. The explored parameter range spans $10^7\leq \textrm{Ra}\leq 10^9$, $1\leq \textrm{Pr}\leq 8$, and $10^{-4}\leq f\leq4$. We are aware of the limitation of the two-dimensionality of the system on which we focus, but in particular for $\textrm{Pr}\geq 1$ two- and three-dimensional RB convection show very close similarities and features \cite{van2013comparison}. To support that our results are also relevant for 3D RB, we conduct a set of three-dimensional DNS in a cubic box at $\textrm{Ra}=10^8$ and $\textrm{Pr}=4.3$ with various frequencies.
\begin{figure}
 \centering
 \includegraphics[width=0.65\columnwidth]{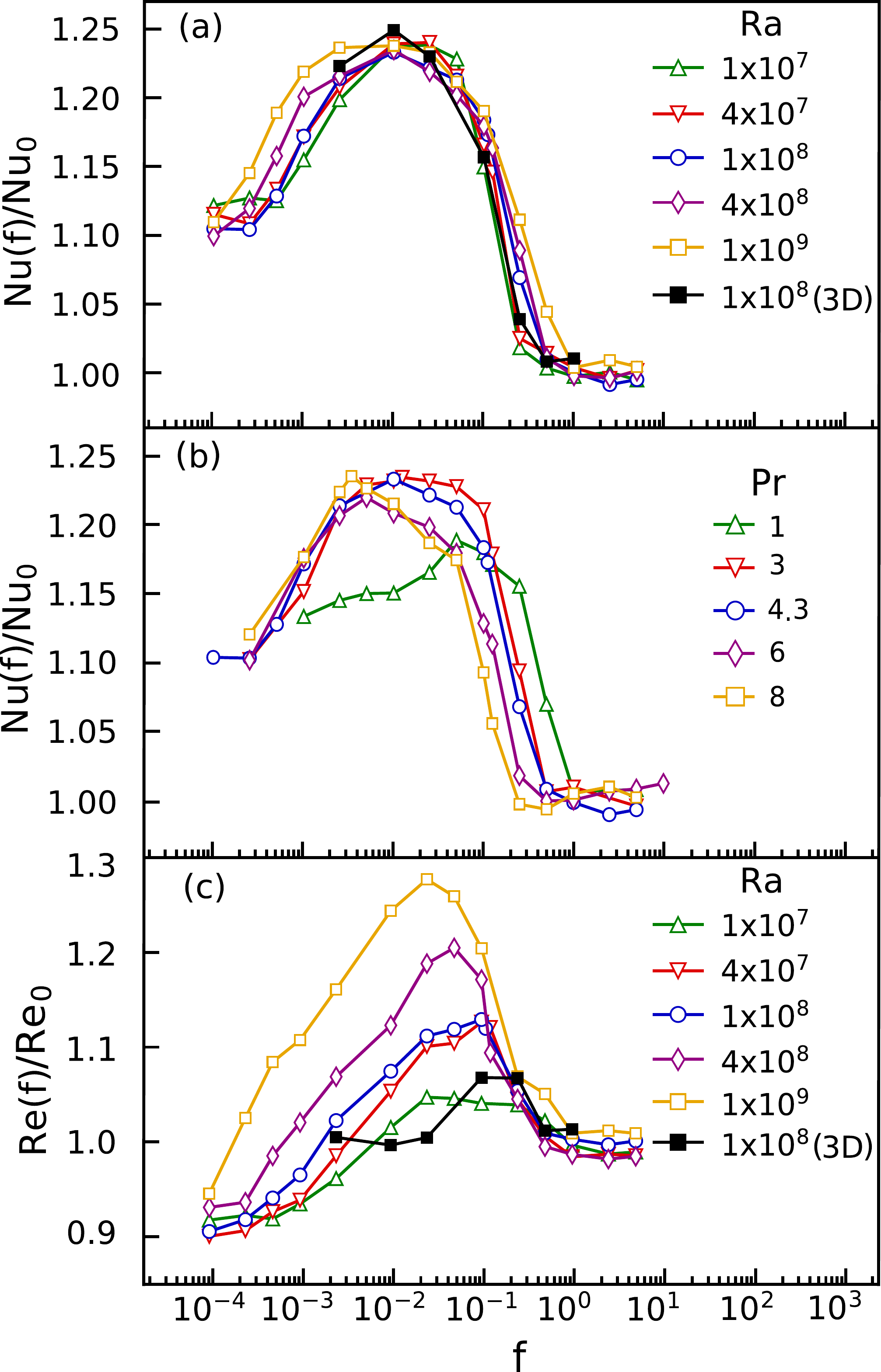}
 \caption{(a) Modulated frequency dependence of the Nusselt number $\textrm{Nu}(f)$, normalized by $\textrm{Nu}_0=\textrm{Nu}(f=0)$, for different Rayleigh numbers and fixed $\textrm{Pr}=4.3$. (b) $\textrm{Nu}(f)/\textrm{Nu}_0$ for different Prandtl numbers and fixed $\textrm{Ra}=10^8$. (c) Global $\textrm{Re}(f)$ normalized by the $\textrm{Re}_0=\textrm{Re}(f=0)$ for different Rayleigh numbers and fixed $\textrm{Pr}=4.3$.}
 \label{fig1}
\end{figure}

Fig.~\ref{fig1}(a) shows how the global convective heat flux $\rm{Nu}$ depends on the modulation frequency $f$ at fixed $\textrm{Pr}=4.3$ (corresponding to water). The dependence of $\rm{Nu}$ on $f$ exhibits a universal trend for both, two- and three-dimensional results, which is independent of $\rm{Ra}$: When $f$ is large enough, $\rm{Nu}$ is not sensitive to the modulation frequency, and the value is close to the value $\textrm{Nu}_0$ for the case without modulation. However, when $f$ is below a certain onset frequency (denoted as $f_\textrm{onset}$), there exists an intermediate regime with significantly enhanced heat flux as compared to $\textrm{Nu}_0$. With $f$ decreasing further, one observes an optimal frequency $f_\textrm{opt}$ at which $\rm{Nu}$ is maximal with an enhancement of approximately $25\%$. Such a large enhancement of $\rm{Nu}$ is highly non-trivial because the time-averaged temperature of the bottom plate is still fixed at $1$, and we only have changed the bottom temperature from a steady value to a time periodic signal. In Fig.~\ref{fig1}(b), we further examine the $\textrm{Nu}(f)$ dependence for different $\textrm{Pr}$, with $\textrm{Ra}$ fixed at $10^8$. One can see that both $f_\textrm{onset}$ and $f_\textrm{opt}$ are much more sensitive to $\textrm{Pr}$ than to $\textrm{Ra}$.
\begin{figure*}
 \centering
 \includegraphics[width=1.5\columnwidth]{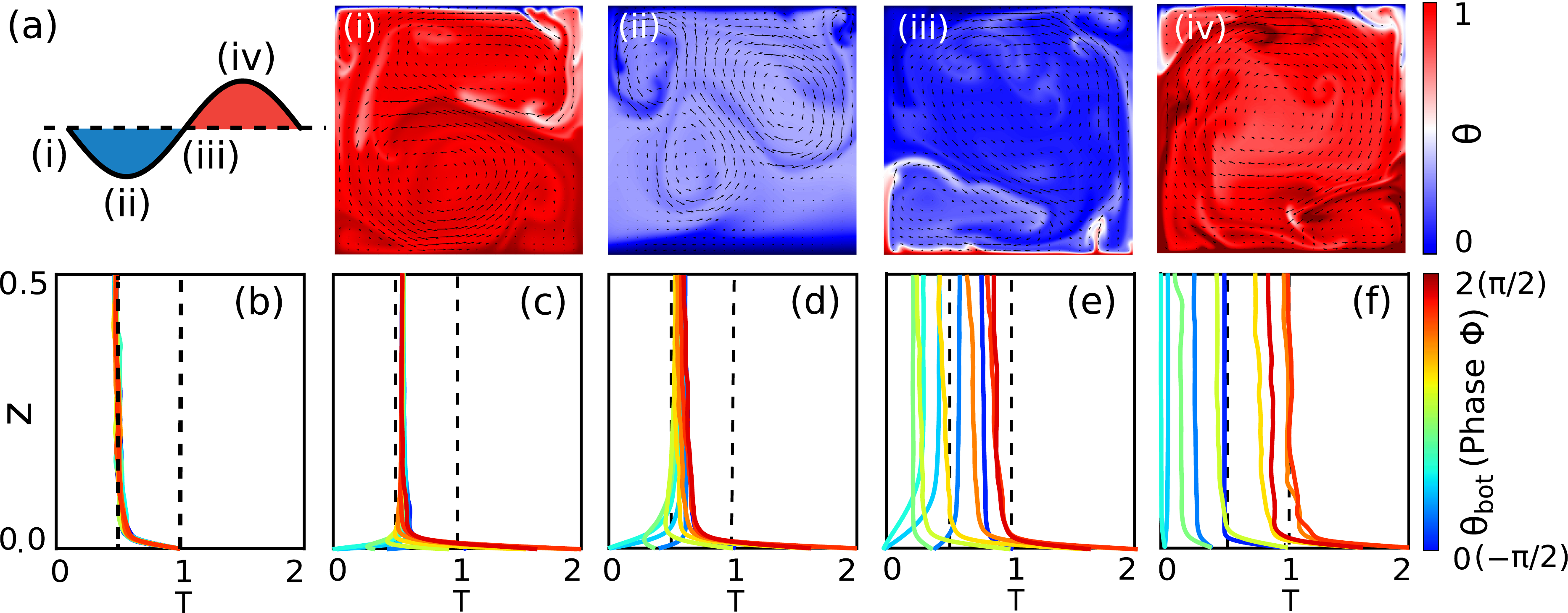}
 \caption{(a) Instantaneous temperature fields at different phases in one modulation period for $\textrm{Ra}=10^8, \textrm{Pr}=4.3, f=10^{-3}$ (see also movies in the Supplemental Material). During the heating phase (iv), the bulk temperature is higher than 0.5 and is shown in reddish color. A strong convective wind and plume emission are observed. In contrast, in the cooling phase (ii), the wind is almost damped out and no plume emission from the bottom can occur due to the stable stratification. (b)-(f) show the phase-averaged temperature profiles during one period for $\textrm{Ra}=10^8, \rm{Pr}=4.3$ and different modulation frequencies, namely (b) without modulation; (c) $f=10^{-1}$; (d) $f=10^{-2}$; (e) $f=10^{-3}$; (f) $f=10^{-4}$. The colorbar shows the bottom temperature (phase angle) from $0(-\rm{\pi}/2)$ to $2(\rm{\pi}/2)$. It is obvious that high-frequency modulation effects are limited in the boundary region, while at low frequency, the system completely follows the modulation as a quasi-steady state.}
 \label{fig2}
\end{figure*}

We first examine whether the transition is related to the strength of the large-scale circulation (LSC). Fig.~\ref{fig1}(c) shows the global Reynolds number $\textrm{Re}$ as function of $f$ for various $\textrm{Ra}$, from which we can see that $\textrm{Re}$ is maximized at a $\textrm{Ra}$-dependent frequency $f_\textrm{opt, Re}$ ( see Reynolds Resonance in Supplemental Material for further analysis of $f_\textrm{opt, Re}$). However, when comparing the $\rm{Nu}$ and $\textrm{Re}$ behaviour, one observes that the position of the strongest LSC does not correspond to that of the maximum heat transport ($f_\textrm{opt}\neq f_\textrm{opt, Re}$). What physics then governs the transitions between the regimes of heat flux?

To gain insight into this problem, we analyze how the flow structure is changed under modulation. Fig.~\ref{fig2}(a) shows the temperature fields at different phases of modulation at $f=10^{-3}$. During the heating phase ($\theta_{bot} > 1$), frequent plume emissions are observed near the bottom plate. On the contrary, during the cooling phase ($\theta_{bot} < 1$), there are no plume emissions from the bottom plate because of the stable stratification near that surface, and the resulting weakening of the circulation. 

We further calculate the conditional average of the temperature profiles at different phases, and compare these profiles for different modulation frequencies in Figs.~\ref{fig2}(b-f). Without modulation, we recover traditional RB with a mean bulk temperature of $0.5$ [Fig.~\ref{fig2}(b)]. When $f=10^{-1}$ as shown in Fig.~\ref{fig2}(c), the temperature adjacent to the bottom is significantly affected by modulation, whereas the bulk value is still close to $0.5$. However, the overall influence of the modulation is limited because it is too fast to be sensed by the system. With decreasing modulation frequency, the bulk temperature is more and more influenced by the modulation (see Fig. 1 of the Supplemental Material). This suggests that there exists a certain length scale which characterizes how deep the influence of the modulation can penetrate into the convective flow.

\begin{figure}[hb]
 \centering
 \includegraphics[width=0.8\columnwidth]{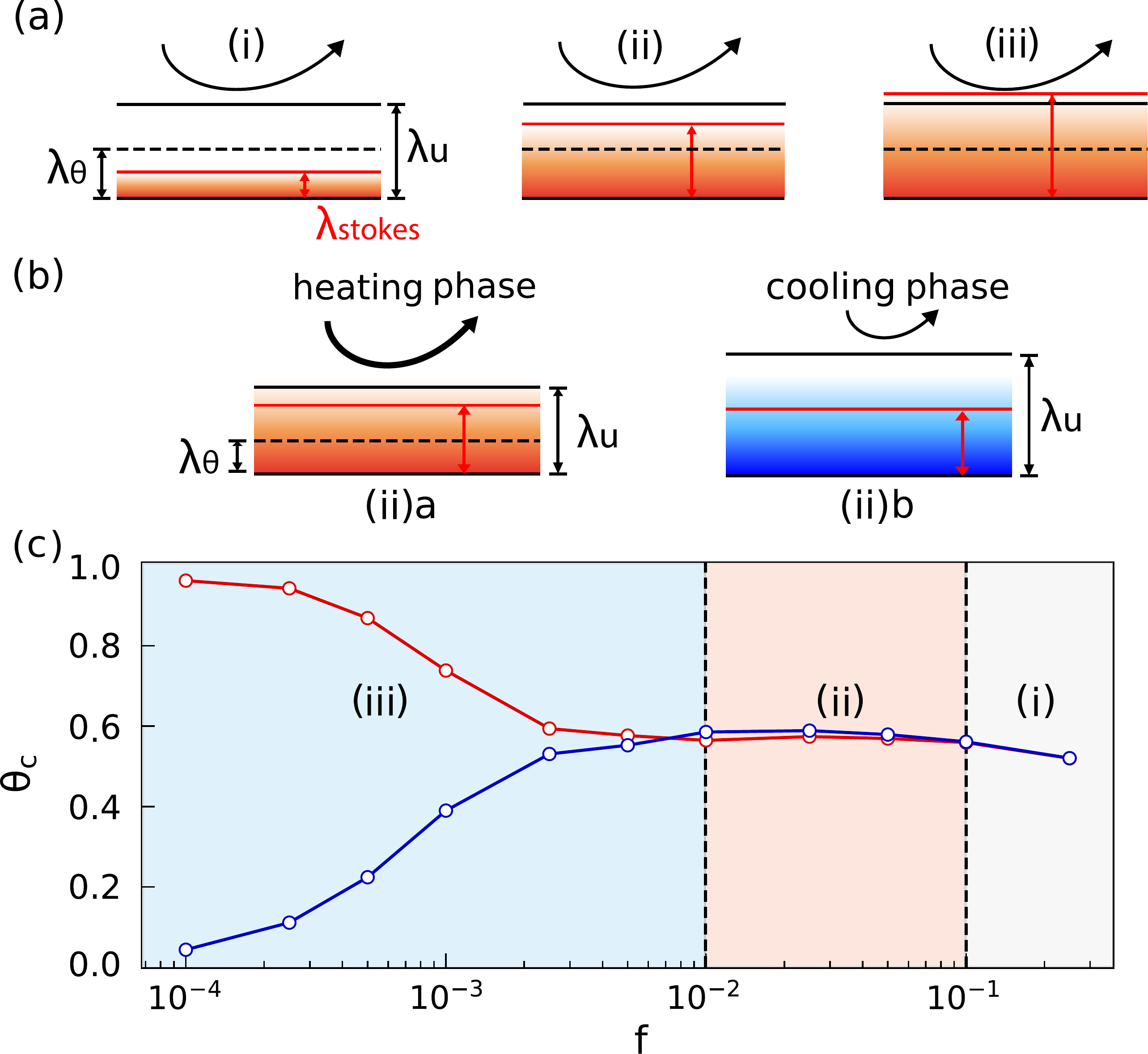}
 \caption{(a) Sketch of the relations between the three BLs (Stokes thermal BL ($\lambda_S$), thermal BL ($\lambda_\theta$), momentum BL ($\lambda_u$)) for the three regimes ($\rm{Pr}=4.3$): (i) $\lambda_u > \lambda_\theta > \lambda_S$; (ii) $\lambda_u > \lambda_S > \lambda_\theta$; (iii) $\lambda_S > \lambda_u > \lambda_\theta$; (b) The sketch of two different phases during one period for regime ii: (a) heating phase when $\theta_{bot}>1$ and (b) cooling phase when $\theta_{bot}<1$; (c) Phase-averaged center temperature for $\textrm{Ra}=10^8, \textrm{Pr}=4.3$. The red (blue) curve represents the phase when the bottom temperature is maximal (minimal). The dash lines (from right to left) correspond to $f_\textrm{onset}$ and $f_\textrm{opt}$ for Nu.
}
 \label{fig3}
\end{figure}

To better understand this length scale, we recall the classical Stokes problem. In this flow, a BL is created by an oscillating solid surface with modulating velocity $ U\cos (2\pi ft)$. Likewise, in modulated RB, we can draw the analogy between an oscillating velocity and the oscillating temperature $\theta^\prime$, where $\theta^\prime=\theta - \bar{\theta}(z)$, with $\bar{\theta}(z)$ being the temporally-averaged temperature at height $z$. The governing equation and corresponding BCs are
\begin{equation}
\begin{split}
\partial \theta^\prime/\partial t &= (RaPr)^{-1/2} \partial^2 \theta^\prime/\partial z^2,\\
\quad \theta^\prime(0,t) &= A\cos (2\pi ft), \quad\theta^\prime(\infty,t) = 0.
\end{split}
\end{equation}
The analytical solution of this PDE is an exponential profile:
\begin{equation}\label{eq:dist}
\theta^\prime(z,t)=A e^{-z/\lambda_S}\cos \left (  2\pi f t-z/\lambda_S \right ),
\end{equation}
with the so-called Stokes thermal BL thickness
\begin{equation}\label{eq:stokes}
\lambda_{S}=\pi^{-1/2}f^{-1/2}\textrm{Ra}^{-1/4}\textrm{Pr}^{-1/4},
\end{equation}
which is the penetration depth of the disturbance created by the oscillating temperature at the boundary. The distortion (Eq. (\ref{eq:dist})) travels as a transverse wave through the fluid. 

From Eq. (\ref{eq:stokes}) one can see that the thickness $\lambda_S$ of the Stokes thermal BL decreases with increasing modulation frequency. Depending on the relative thicknesses of $\lambda_S$, that of the thermal BL $\lambda_\theta$, and that of the momentum BL $\lambda_u$, we can obtain three regimes shown in Fig.~\ref{fig3}(a). Here we have restricted us to $1\leq \textrm{Pr}\leq 8$, where $\lambda_u\geq\lambda_\theta$.

\begin{figure*}[ht]
 \centering
 \includegraphics[width=1.8\columnwidth]{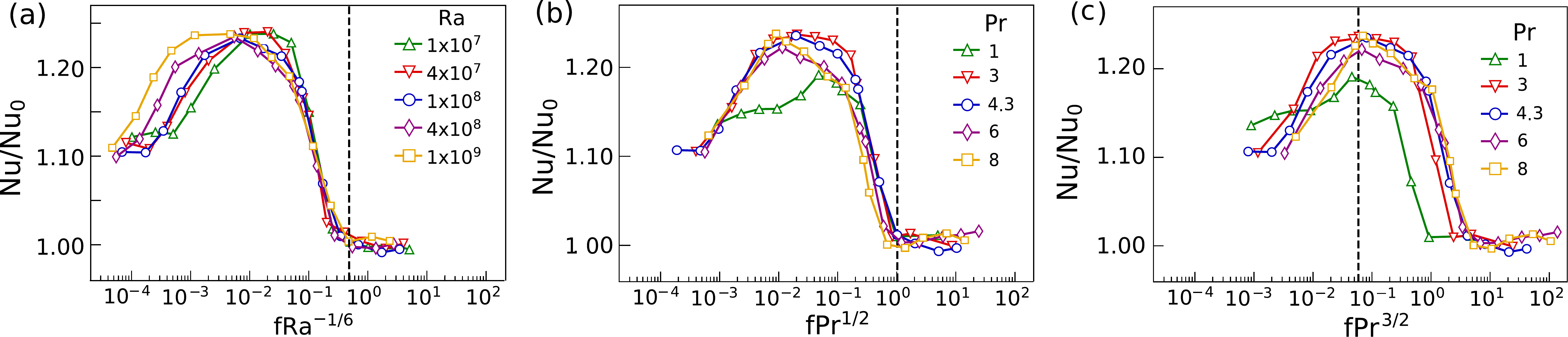}
 \caption{(a) The normalized $\rm{Nu}$ as function of $fRa^{-1/6}$, for different Ra and $\rm{Pr}=4.3$, (b) $fPr^{1/2}$, (c) $fPr^{3/2}$ for different $\rm{Pr}$ and $\rm{Ra}=10^8$. The dash lines show the onset frequency (where $\textrm{Nu}(f)$ starts to be affected, $\textrm{Nu}(f)/\textrm{Nu}_0 = 1.01$) or optimal frequency (where $\textrm{Nu}(f)$ reaches the maximum), averaged for different $\rm{Ra}$ or $\rm{Pr}$.}
 \label{fig4}
\end{figure*}

\begin{itemize}[leftmargin=*]

\item Regime (i): For $\lambda_S<\lambda_\theta<\lambda_u$, the effect of modulation is confined inside the thermal BL, which is also shown by the temperature profiles in Fig.~\ref{fig2}(c). In such case, the effect of modulation is negligible and the heat transport is almost unaffected. 

\item Regime (ii): For $\lambda_\theta\leq\lambda_S<\lambda_u$, the plume emission, which occurs at the edge of the thermal BL, can now be influenced by the modulation [Fig.~\ref{fig3}(b)]. Subsequently, the enhancement in $\rm{Nu}$ can be accounted for by the following mechanism: In the heating phase ($\theta_{bot}>1$), stronger convection and more frequent plume emission are produced by a hotter bottom plate, as compared to the case without modulation. However, even when $\theta_{bot}=0$ in the cooling phase, the convective flow is not completely damped out due to the thermal inertia, which therefore enhances the heat flux as compared to the pure diffusion case. With the remaining convective flow which enhances the heat transport in cooling phase, there is a net increase in $\rm{Nu}$ after one cycle compared to the $\rm{Nu}$ without time-dependent modulation.

\item Regime (iii): For $\lambda_\theta<\lambda_u\leq\lambda_S$, the effect of temperature modulation penetrates into the bulk region occupied by the LSC. In this case, the bulk temperature becomes sensitive to modulation. As seen from Fig.~\ref{fig3}(c), the bulk temperatures during the cooling and heating phases indeed deviate from each other when $f<10^{-2}$, i.e., below the optimal frequency. As a result, at the peak of the heating phase ($\theta_{bot}=2$), the temperature difference between the bottom plate and the bulk cannot be maintained at $\Delta\theta\simeq1.5$. The thermal driving in the heating phase becomes weaker for smaller $f$, and the global $\rm{Nu}$ is expected to decrease for decreasing $f$. 
\end{itemize}

To obtain the scaling laws for $f_{\textrm{onset}} (\textrm{Ra},\textrm{Pr})$ and $f_{\rm{opt}}(\textrm{Ra},\textrm{Pr})$, we compare the BL thicknesses. First, we make use of the relations $\lambda_\theta \sim \textrm{Nu}^{-1}$ and $\lambda_u \sim \textrm{Re}^{-1/2}$ for the thermal and momentum BL thicknesses. Then we use the Grossmann-Lohse model for the scaling of $\textrm{Nu}(\textrm{Ra},\textrm{Pr})$ and $\textrm{Re}(\textrm{Ra},\textrm{Pr})$ in the $\textrm{I}_\infty$ regime (for large Pr) \cite{Grossmann2000, Shishkina2017}: $\textrm{Nu}\sim \textrm{Pr}^0\textrm{Ra}^{1/3}$ and $\textrm{Re} \sim \textrm{Pr}^{-1} \textrm{Ra}^{2/3}$. The onset frequency $f_\textrm{onset}$ corresponds to the transition between Regime i and Regime ii ($\lambda_S\sim\lambda_\theta$), and we obtain
\begin{equation}
\label{eq:fon}
    f_{\rm{onset}} \sim\rm{Ra}^{1/6}\rm{Pr}^{-1/2}.
\end{equation}
The optimal frequency $f_\textrm{opt}$ corresponds to the transition between Regime ii and Regime iii ($\lambda_S\sim\lambda_u$), and we have
\begin{equation}
\label{eq:fopt}
    f_{\rm{opt}} \sim \rm{Ra}^{1/6}\rm{Pr}^{-3/2}.
\end{equation}
\begin{figure}[hb]
 \centering
 \includegraphics[width=0.9\columnwidth]{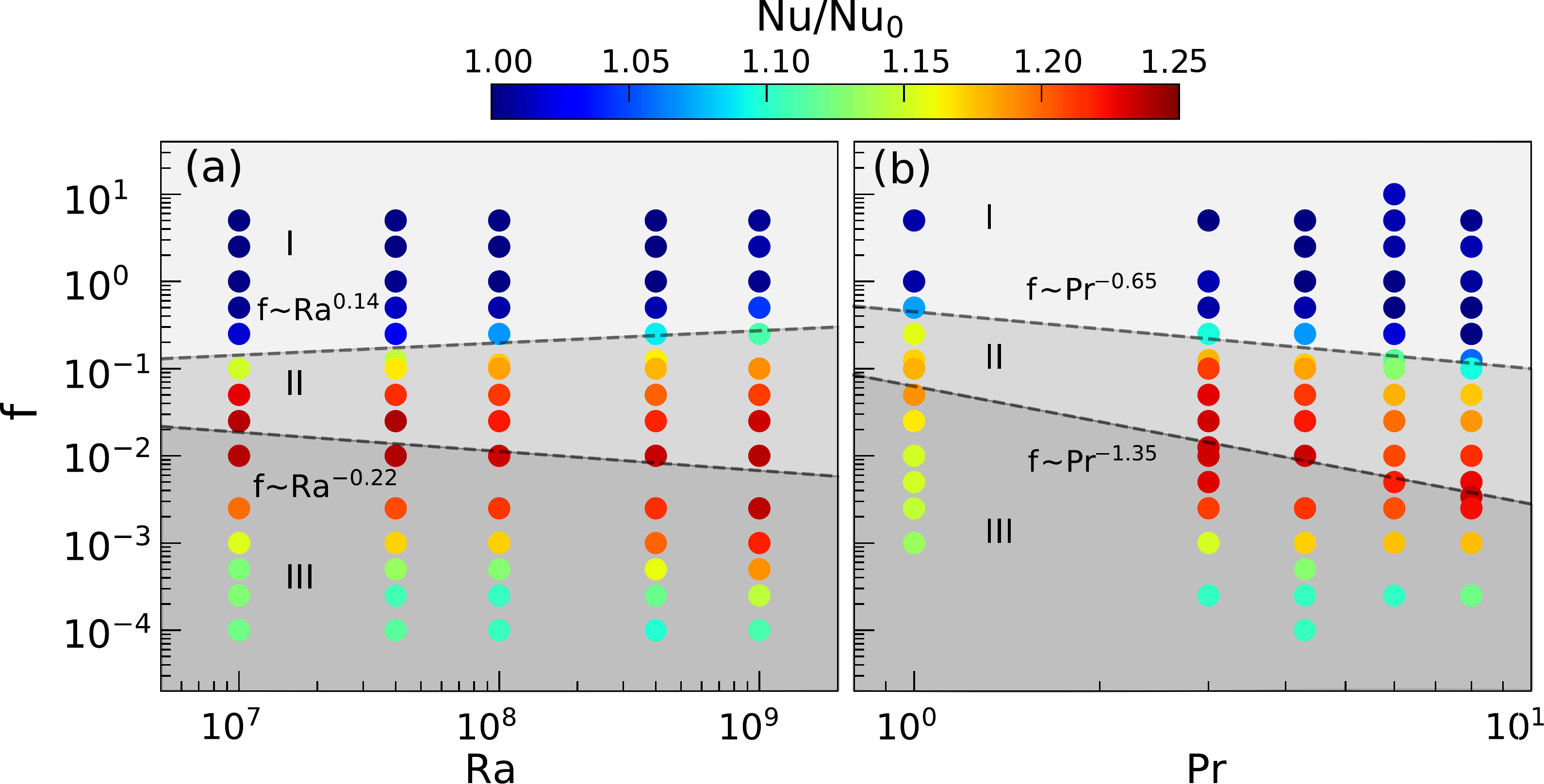}
 \caption{Phase diagram (a) in the $f$ vs. $\textrm{Ra}$ and (b) in the $f$ vs. $\textrm{Pr}$ parameter spaces. In (a), the lower dashed line shows the optimal frequency $f_{opt} = 0.65\textrm{Ra}^{-0.22}$ that corresponds to the maximal $\rm{Nu}$. The upper dashed line shows the onset frequency $f_{onset} = 0.015\textrm{Ra}^{0.14}$ that corresponds to the onset of the heat flux enhancement. In (b), the lower dashed line shows the optimal frequency $f_\textrm{opt} = 0.06\textrm{Pr}^{-1.35}$, while the upper one shows the onset frequency $f_\textrm{onset} = 0.45\textrm{Pr}^{-0.65}$. The prefactors originate from fits to the DNS data for $f_{opt}$ and $f_{onset}$ (set to occur when $\textrm{Nu}(f)/\textrm{Nu}_0=1.01$), see Supplemental Material for details on the fitting.}
 \label{fig5}
\end{figure}

To check these predictions for $f_{\rm{onset}}$ and $f_{\rm{opt}}$, we replot $\textrm{Nu}(f)$ for various $\textrm{Ra}$ but now versus the \textit{rescaled} frequency $f\textrm{Ra}^{-1/6}$, see Fig.~\ref{fig4}(a) ($\textrm{Pr}=4.3$ fixed). Indeed, the figure shows rather good collapses around the onset. Next, we vary $\textrm{Pr}$ for a fixed $\textrm{Ra}=10^8$ and plot $\textrm{Nu}$ versus the correspondingly rescaled frequencies, namely $f\textrm{Pr}^{1/2}$ for the onset [Fig.~\ref{fig4}(b)] and $f\textrm{Pr}^{3/2}$ for the optimum [Fig.~\ref{fig4}(c)]. Indeed, one can see that equations (\ref{eq:fon}) and (\ref{eq:fopt}) also correctly predict the onset frequency and the optimal frequency for all $\rm{Pr}$.

Finally, we present the phase diagram in the $f$ vs. $\textrm{Ra}$ and the $f$ vs. $\textrm{Pr}$ parameter spaces in Fig.~\ref{fig5}. We classify three regimes: classical RB regime (i), modulation-enhancement regime (ii), and modulation-reduction regime (iii). The boundary between the regimes is found by fitting the numerically obtained $f_\textrm{onset}$ and $f_\textrm{opt}$. The fitting scaling relations for onset and optimum ($f_{\textrm{onset}} \sim\rm{Ra}^{0.14}\rm{Pr}^{-0.65}, f_{\rm{opt}} \sim \rm{Ra}^{-0.22}\rm{Pr}^{-1.35}$) show a good agreement with the derived ones ($f_{\textrm{onset}} \sim\rm{Ra}^{1/6}\rm{Pr}^{-1/2}, f_{\textrm{opt}} \sim \rm{Ra}^{1/6}\rm{Pr}^{-3/2}$) except $f_\textrm{opt}$ vs. $\rm{Ra}$, corresponding to $\lambda_S\sim\lambda_u$. We notice that in our model, $\lambda_S$ is obtained based on diffusion equation. The neglected advection term can become significant, particularly in regime iii where the Stokes BL may penetrate into the bulk. It, therefore, imposes uncertainty in estimating the weak Ra-dependence of $\lambda_\textrm{opt}$.

In conclusion, our results have substantial implications for the investigation of modulated convection systems. For a wide range of parameter in the three-dimensional parameter space (modulation frequency $f$, Rayleigh number $\rm{Ra}$ and Prandtl number $\rm{Pr}$), we have demonstrated how the global heat transport efficiency can be enhanced through temperature modulation in both two- and three-dimensional simulations. The high similarity between 2D and 3D DNS results supports that our results are applicable in both cases and robust. Based on the heat transfer enhancement, we can identify three different regimes: the classical RB regime for fast modulation, an intermediate regime in which the modulation leads to increasing Nu-enhancement, and the slow modulation regime in which it leads to decreasing Nu-enhancement. The transitions between the regimes are well predicted by the relative thicknesses of thermal, momentum and Stokes thermal BLs. Our concept of explaining global transport properties in modulated BL flows by the relative thicknesses of the three relevant BLs can also be extended to the angular velocity transfer in modulated turbulent Taylor-Couette flow, or to the kinetic energy transfers in modulated turbulent pipe flow.

\textit{Acknowledgement} This work was supported by the Priority Programme SPP 1881 Turbulent Superstructures of the Deutsche Forschungsgemeinschaft and by NWO via the Zwaartekrachtprogramma MCEC and an ERC-Advanced Grant under the project number 740479. This work was partly carried out on the national e-infrastructure of SURFsara,

R. Y. and K. L. C. contributed equally to this work.

\end{document}